# Dielectric Relaxation in Nanopillar NiFe-Silicon Structures in High Magnetic Fields

R. Vasic[a], J.S. Brooks[a], E. Jobiliong[a], S.Aravamudhan[b], K.Luongo[b], and S.Bhansali[b]

[a]*Department of Physics and National High Magnetic Field Laboratory, Florida State University, Tallahassee, Florida 32306*

[b]*Department of Electrical Engineering, University of South Florida, 4202 E Fowler Ave, Tampa, 33620 State University,*

**Abstract**

We explore the dielectric relaxation properties of NiFe nanowires in a nanoporous silicon template. Dielectric data of the NiFe-silicon structure show a strong relaxation resonance near 30K. This system shows Arrhenius type of behavior in the temperature dependence of dissipation peaks vs. frequency. We report magnetic field dependence of dipolar relaxation rate and the appearance of structure in the dielectric spectrum related to multiple relaxation rates. A magnetic field affects both the exponential prefactor in the Arrhenius formula and the activation energy. From this field dependence we derive a simple exponential field dependence for the prefactor and linear field approximation for the activation energy which describes the data. We find a significant angular dependence of the dielectric relaxation spectrum for regular silicon and nanostructured silicon vs. magnetic field direction, and describe a simple sum-rule that describes this dependence. We find that although similar behaviour is observed in both template and nanostructured materials, the NiFe-silicon shows a more complex, magnetic field dependent relaxation spectrum.

## 1. Introduction

Dielectric relaxation is an important method to describe insulating or semiconducting systems since dc transport on very high resistance samples becomes increasingly difficult at low temperatures. Whenever the system in the insulating state exhibits a dipolar structure it is possible to find dielectric relaxation in some range of temperature and frequency due to the resonance condition $\ln(f) \sim 1/T$. In the present work we have been exploring the dielectric relaxation of NiFe nanowires electrodeposited in nanoporous silicon templates[1] as shown in Fig. 1. Although our work follows the general dielectric response seen in for semiconducting materials[2-6], the present investigation extends such studies to very high magnetic fields, where multiple relaxation structure emerges, and magnetic field direction dependent dielectric relaxation appears.

## 2. Experimental methods and results

Fabrication of magnetic nanowire arrays by electrodeposition is important for diverse applications in fields such as magnetic recording and bio-magnetics.[7-9] The nanoporous silicon was prepared by electrochemical etching of silicon substrate in a sulphate based electroplating bath[1]. The target pore diameter was controlled by resistance of template and the length of nanopores by changing of the etching time. Later NiFe was electrochemicly deposited from a sulphate based electroplating bath into nanoporous silicon by using a Ni foil as anode. Fig. 1 shows a scanning electron microscope(SEM) image of the structure and morphology of the nanostructured Si template. [1] Parameters associated with these materials include the electrical transport gap of 0.153eV, which makes the dc conductivity difficult to measure below about 50 K. The NiFe ratio is 81:19, i.e. permalloy. The coercive field is below 500 gauss, and the magnetization exhibits a Curie law dependence with decreasing temperature. There is no evidence for magnetic order above 4.2 K.

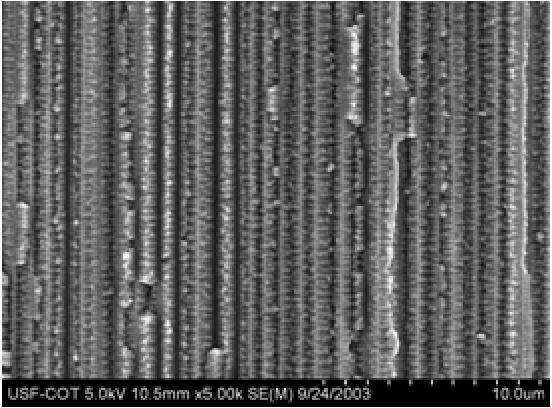

Figure 1. High resolution SEM image of 290 nm diameter and 145 μm deep nanopores electrochemically etched in n-type Si substrate (resistivity: 0.4-0.6 Ohm-cm).

Dielectric measurements were carried out on [100] oriented Si:P and nanopillar NiFe-silicon. Sample geometries were rectangular, with dimensions typically 2 mm × 1 mm × 0.5 mm. For both samples results are collected with parallel plate silver paste electrodes normal to [100] direction, and in the case of the NiFe-Si, also normal to the nanopillars. The real (capacitative - C) and loss (dissipative - D) signals were measured with an ac capacitance bridge and lock-in amplifier. Measurements were carried out vs. temperature, frequency, and magnetic field at the National High Magnetic Field Laboratory. The results presented in Figs. 2-4 are obtained for ac electrical field perpendicular to the direction of the magnetic field (90° in the present notation).

A dielectric system with a characteristic dipolar relaxation time $\tau = \varepsilon/\sigma$ has a characteristic frequency and temperature where the response exhibits a peak in the dissipation D (or $\varepsilon''$) and a phase shift in the capacitance C (or $\varepsilon'$) response.[10-12] The relationship between f and T at resonance will follow an Arrhenius form for the relaxation rate $1/\tau = f(T) = f_0 \exp(-E_a/T)$. Typical results for the frequency dependence are shown in Fig. 2 for the nanostructured NiFe silicon system where capacitance and dissipation are monitored vs. temperature and frequency for zero magnetic field and in Fig. 3 for 20 T. In Fig. 4 the magnetic field dependence is shown for a frequency of 100 kHz. We find that the general behavior seen in Figs. 2-4 is characteristic of both the silicon template which is a Si:P system with about $10^{14}$ carriers/cm$^3$, and for the NiFe system, except for some important differences to be discussed in Sec. 3 below. In what follows, our analysis is carried out on the most pronounced peak feature in the dissipation.

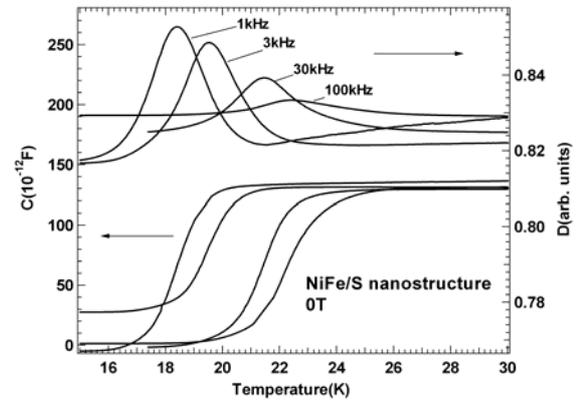

Figure 2. Temperature dependence of real (left axis) and imaginary (right axis) dielectric constant for 1, 3, 30, and 100 kHz at zero magnetic field.

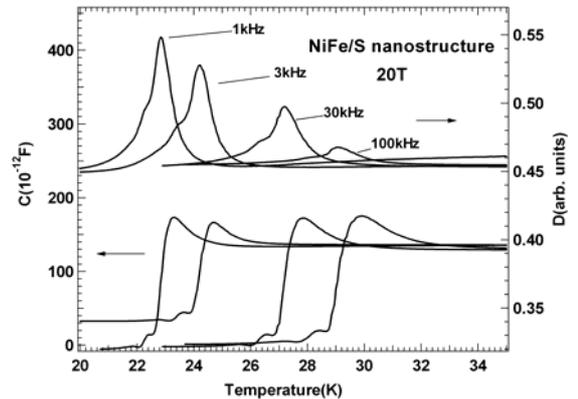

Figure 3. Temperature dependence of real (left axis) and imaginary (right axis) dielectric constant for 1, 3, 30, and 100 kHz at constant magnetic field (20T).

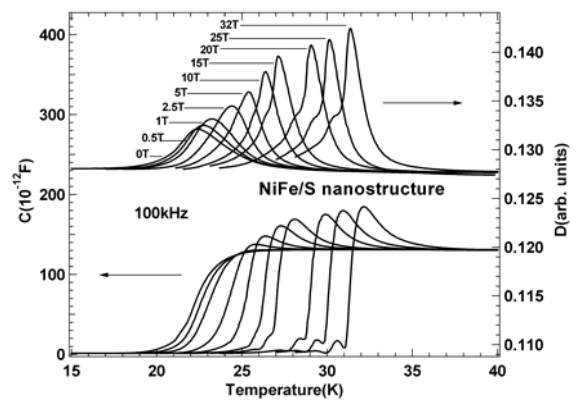

Figure 4. Temperature dependence of real (left axis) and imaginary (right axis) dielectric constant (100 kHz) for different magnetic fields.

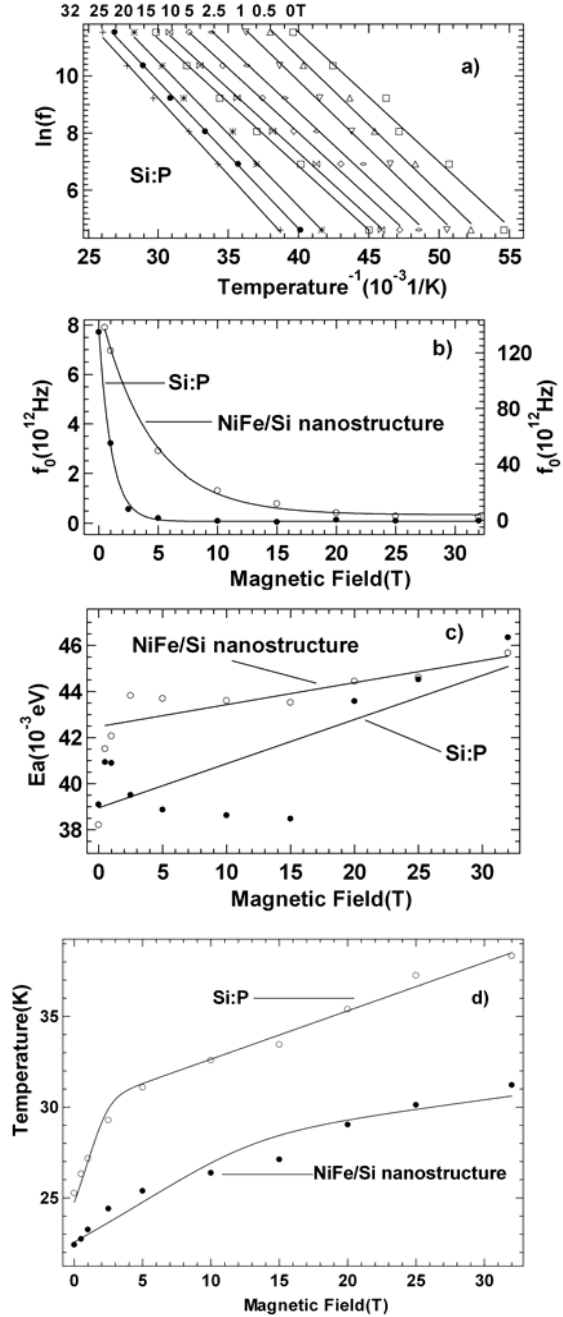

Figure 5. a) Arrhenius plot of relaxation frequencies of silicon for different magnetic fields; lines are fits to Eq.1. (Note: NiFe shows similar behavior); b) Magnetic field dependence of intercept frequency $f_0$ for silicon and NiFe-Si; lines are fits to Eq. 2. c) Magnetic field dependence of activation energy $E_a$; lines are fits to Eq. 3. d) Magnetic field dependence of relaxation peak temperatures at 100 kHz (symbols) and corresponding fits from Eq. 4.

The general frequency-temperature-magnetic field dependence of the dielectric relaxation may be summarized as shown in Fig. 5a. Here, for a specific field, an Arrhenius relation is followed for ln(f) vs. 1/T, where the slope (activation energy $E_a$) and the intercept (prefactor $f_0$ for 1/T ->0) are magnetic field dependent. Fits to the data in Fig. 5a for different fields lead to a description of the field dependence of $f_0$ and $E_a$ as shown in Figs. 5b and 5c respectively, for both Si:P and NiFe-Si. We may therefore parameterize the general behavior of the field dependent dielectric relaxation in the form

$$f(T,B) = f_0(B)\exp(-E_a(B)/T), \quad (1)$$

where

$$f_0(B) = f_1 + f_2\exp(-B/B_0), \quad (2)$$

and

$$E_a(B) = \alpha + \beta B. \quad (3)$$

Parameters $f_1$, $f_2$, and $B_0$ for Si:P and the NiFe-Si system, which exhibit a strong exponential decrease with increasing field, are presented in Table 1.

Table 1. Parameters of the field dependent prefactor for Si:P and NiFe-Si

| $f_0 = f_1 + f_2\exp(-B/B_0)$ | $f_1$(Hz) | $f_2$(Hz) | $B_0$(T) |
|---|---|---|---|
| Si:P | $8.2 \times 10^{10}$ | $7.7 \times 10^{12}$ | 1.06 |
| NiFe silicon | $4.0 \times 10^{12}$ | $1.5 \times 10^{14}$ | 4.3 |

The activation energy has a weak, positive field dependence with large uncertainties, which we have approximated by a linear function of field. Fits to Eq. 3 yield for Si:P, $\alpha = 0.039$eV; $\beta = 0.00019$eV/T; and for nanostructured NiFe silicon, $\alpha = 0.042$eV; $\beta = 0.00014$eV/T.

We may independently test the applicability of Eq. 1 by considering data in the form of the field dependence of the dissipative peaks at constant frequency, as shown in Fig. 4, and plotted in Fig. 5d. In the data we see that the resonant temperature first increases quickly, and then falls to a linear dependence at higher fields. That this follows from the parameters above can be seen by solving Eq. 1 for temperature:

$$T = E_a/\ln(f_0/f). \quad (4)$$

Here, for increasing field at constant frequency, T will increase rapidly as $f_0$ decreases (rapidly) at low fields. At high fields, $f_0$ saturates, but $E_a$ monotonically increases.

We have also explored the effects of magnetic field direction with respect of electric field on the dielectric relaxation in both the template and NiFe processed materials in high fields where the relaxation spectrum shows evidence for multiple dipolar sites. As shown in Figs. 6 and 7, there is a strong dependence of the relaxation spectrum with field direction, indicating that the different dipolar sites are anisotropic. It is interesting to note however, that there is a "sum rule" obeyed, as can be seen by integrating the spectra vs. temperature: i.e., the area under the curves, regardless of field orientation, is essentially constant. This result is represented for both materials in Fig. 8.

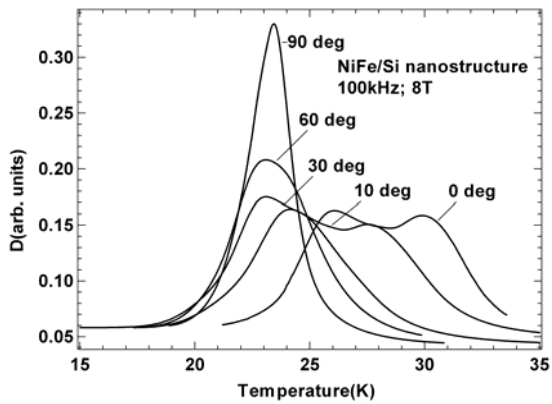

Figure 6. Angular magnetic field dependence of the dielectric relaxation (dissipation) of the NiFe nanostructure at 8 T. (Note: 0°: E∥B ; 90°: E⊥B)

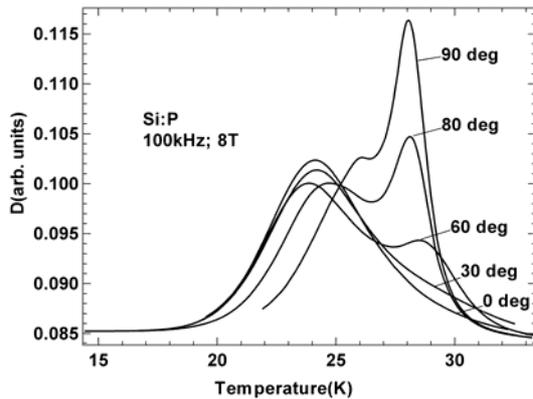

Figure 7. Angular magnetic field dependence of the dielectric relaxation (dissipation) of Si:P at 8 T. (Note: 0°: E∥B ; 90°: E⊥B)

We note that we have independently checked the anisotropy of the relaxation spectrum with field direction on samples with no metal-semiconductor interfaces (i.e., insulating capacitive electrodes) and have confirmed that anisotropy related to planar Schottky barriers at the electrode contacts (without insulation) is not involved.

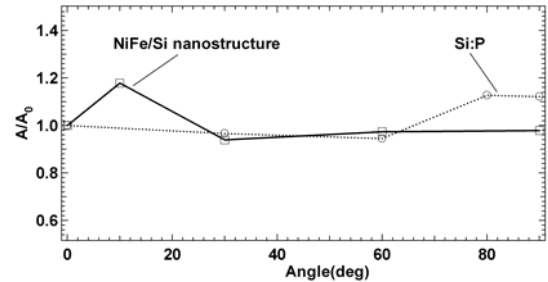

Figure 8. Integrated spectra (from Figs. 6 and 7) for different field orientations normalized to the zero angle values.

## 3. Discussion

Our overall magnetic field dependent dipolar relaxation results are consistent with concept of magnetic localization and magnetic freezeout in semiconductor systems. Magnetic localization develops as the magnetic field shrinks the overlap between phosphorous donors or other impurity sites, making the dipolar charge distribution "stiffer".[5, 13] When the cyclotron radius of electron orbital becomes comparable to the Bohr radius, this induces magnetic freezeout.[5] The effect of magnetic freezeout is to increase the energy barrier between donors because of increased localization of conduction electrons on the donor sites. In light of this, have identified two processes: the attempt frequency in the Arrhenius law is suppressed with increasing field (i.e. the dipolar response slows down), and second, the energy gap increases (i.e. the inter-donor barrier increases).

We find that the dependence of the dipolar relaxation rate on magnetic field and temperature is a general feature of n-type silicon (n ~ $10^{14}$/cm$^3$) and the NiFe- Si systems. This includes the appropriateness of simple Debye theory both for dissipation and capacitance components, where the intensity of the imaginary part of dielectric function decreases with increasing frequency. Likewise, the dissipation peaks increase and the linewidths decrease significantly with higher magnetic fields. Finally, in both systems, higher magnetic fields (generally above about 5 T) produce multiple structure in the relaxation spectrum that indicate additional dipolar sites, even in the less complicated Si:P material.

There are however, several important differences in the magnetic field dependent relaxation spectra between the template and nanostructured materials. We find that in the NiFe-Si system there are there are additional components that lead to multiple, field dependent dipolar rates. In the X-ray spectroscopic analysis, the presence of Ni, NiSi, $NiSiO_4$ and $Ni_3Fe$ alloy reflections appear, indicating additional non-phosphorous sites. A broader distribution of sites is consistent with our observation of a broader cutoff field $B_0$ and larger frequency pre-factor in the NiFe system. It is not clear at this point if the actual nano-structure of the NiFe pillars is influential in the dielectric relaxation beyond the contribution of additional dipolar structures. One anomaly in the NiFe-Si system that is not explained, based on a purely Debye model is the peak in the capacitive signal just above the resonant temperature (Fig. 4c). Likewise, the anisotropy in the Si:P and NiFe-Si systems appears to be reversed, since in the former, a single peak appears for E//B, and in the latter the single peak appears for E⊥B.

## 4. Conclusions

The main finding of our work is that the dipolar relaxation in lightly doped silicon and nanostructured silicon exhibit a pronounced, anisotropic magnetic field dependence. In the case of the nanostructured materials, high magnetic fields reveal additional dipolar sites due to the more complex composition of the materials. We may model this behavior in terms of a field dependent "attempt frequency" and activation energy associated with the dipolar relaxation. Moreover, we find that a sum rule governs the integrated dielectric relaxation spectrum regardless of magnetic field direction. More studies are underway to explain mechanism of dielectric relaxation in other nanostructured systems and novel dielectric materials.

## Acknowledgements

Supported by NSF-DMR 02-03532. NHMFL is supported by the NSF and the State of Florida.

## 3. References

[1]. S. Aravamudhan, K. Luongo, S. Bhansali, P. Poddar, H. Srikanth, Metallurgical and Materials Transactions A.(to be published).
[2] T. W. Hickmott, Phys. Rev. B **32,** (1985) 6531
[3] T. W. Hickmott, Phys. Rev. B **38,** (1988) 12404
[4] T. W. Hickmott, Phys. Rev. B **44,** (1991) 13487
[5] T. W. Hickmott, Phys. Rev. B **46,** (1992) 12342
[6] S. Aymeloglu, J. N. Zemel, IEEE Trans. Electron. Dev. **ED-23**, (1976) 466.
[7] A. Fert, L. Piraux, J. Magn. Magn. Mater. **200,** (1999) 338-358.
[8] I. Safarik, M. Safarikova, J. Chromatogr. **722,** (1999)33.
[9] D. H. Reich, M. Tanase, A. Hultgren, L. A. Bauer, C. S. Chen, G. J. Meyer, J. Appl. Phys. **93,** (2003) 7275.
[10] P. Debye, Phys. Z. **13** (1912) 97.
[11] N. E. Hill, W. E. Vaughn, A. H. Price, and M. Davies, *Dielectric Properties and Molecular Behavior* (Van Nostrand Reinhold, London,1969).
[12] K. S. Cole and R. H. Cole, J. Chem. Phys. **9,** (1941**)** 341.
[13] Y. Yafet, W. Keyes, and E. N. Adams, J. Phys. Chem. Solids **1,** (1956) 137.